\documentclass[letter]{aa}
\usepackage{graphicx}
\usepackage{txfonts}
\input epsf

\begin{document}

\title{Mode lifetimes of stellar oscillations}

\subtitle{Implications for asteroseismology}

\author{W.~J. Chaplin \inst{1}, G. Houdek \inst{2,3}, C. Karoff
\inst{1}, Y. Elsworth \inst{1} \and R. New \inst{4}}

\institute{School of Physics and Astronomy, University of Birmingham,
  Edgbaston, Birmingham B15 2TT, UK\\ \email{w.j.chaplin@bham.ac.uk} \and
  Institut f\"ur Astronomie, Universit\"at Wien, T\"urkenschanzstra{\ss}e 17, 1180 Vienna, Austria \and
  Institute of Astronomy, University of Cambridge, Cambridge CB3 0HA,
  UK \and Faculty of Arts, Computing, Engineering and Sciences,
  Sheffield Hallam University, Sheffield S1 1WB, UK}

   \date{Received ; accepted }
 
  \abstract {Successful inference from asteroseismology relies on at
    least two factors: that the oscillations in the stars have
    amplitudes large enough to be clearly observable, and that the
    oscillations themselves be stable enough to enable precise
    measurements of mode frequencies and other parameters. Solar-like
    p modes are damped by convection, and hence the stability of the
    modes depends on the lifetime.} {We seek a simple scaling relation
    between the mean lifetime of the most prominent solar-like p modes
    in stars, and the fundamental stellar parameters.} {We base our
    search for a relation on the use of stellar equilibrium and
    pulsation computations of a grid of stellar models, and the first
    asteroseismic results on lifetimes of main-sequence, sub-giant and
    red-giant stars.} {We find that the mean lifetimes of all three
    classes of solar-like stars scale like $T_{\rm eff}^{-4}$ (where
    $T_{\rm eff}$ is the effective temperature). When this relation is
    combined with the well-known scaling relation of Kjeldsen \&
    Bedding for mode amplitudes observed in narrow-band intensity
    observations, we obtain the unexpected result that the height (the
    maximum power spectral density) of mode peaks in the frequency
    power spectrum scales as $g^{-2}$ (where $g$ is the surface
    gravity). As it is the mode height (and not the amplitude) that
    fixes the S/N at which the modes can be measured, and as $g$
    changes only slowly along the main sequence, this suggests that
    stars cooler than the Sun might be as good targets for
    asteroseismology as their hotter counterparts. When observations
    are instead made in Doppler velocity, our results imply that mode
    height does increase with increasing effective temperature.

}{}


   \maketitle

\section{Introduction}
\label{sec:intro}

Asteroseismology is now opening new windows on the interiors of
solar-like stars, and allowing inferences to be made on the masses,
sizes, and ages to levels of precision that would not otherwise be
possible.  These results have come from ground-based campaigns
(Bedding \& Kjeldsen 2006) and from spacecraft observations, e.g.,
WIRE (Bruntt 2007), MOST (Matthews et al. 2007) and SMEI (Tarrant et
al. 2007).  CoRoT (Baglin et al. 2006) is now producing exciting
results on hot solar-like main-sequence stars (Michel et al. 2008);
while NASA's Kepler mission (Basri et al. 2005) will increase by more
than two orders of magnitude the number of solar-like stars we can
observe by seismology (Christensen-Dalsgaard et al. 2007).

A first detailed attempt to predict amplitudes and hence the
detectability of solar-like p-mode oscillations in stars other than
the Sun was presented by Christensen-Dalsgaard \& Frandsen
(1983). They made computations for a grid of stellar models in the
colour-magnitude diagram, and concluded that both velocity and
intensity oscillation amplitudes generally increased with increasing
age and with increasing mass along the main sequence. Kjeldsen \&
Bedding (1995) used the same model data as Christensen-Dalsgaard \&
Frandsen, and found that the p-mode velocity amplitudes in the models
scaled as $v_{\rm osc} \propto \left(L/M\right)^s$ where
$s=1$. Kjeldsen \& Bedding (1995) also used linear theory and
observational data to obtain a relation between the velocity and
intensity amplitudes, i.e., $(\delta L/L)_{\lambda} \propto v_{\rm
osc} / T_{\rm eff}^2$ (assuming the amplitudes in intensity were
inferred from narrow-band observations).

The scaling relations of Kjeldsen \& Bedding (1995) have been used
extensively within the field of asteroseismology. However, several
studies have since suggested modifications to the relations. Houdek et
al. (1999) calculated a new grid of stellar models, and included the
perturbations to the convective heat and momentum fluxes in the mode
stability computations. From this improved grid of stellar models
Houdek et al. (1999) still obtained the same scaling relation as
Kjeldsen \& Bedding (1995) for $L/M$ less than three times the solar
value, but found that $s=1.29$ for $M/M_{\sun} \gtrsim 1.4$. Samadi et
al. (2005), on the other hand, used numerical 3D simulations of
near-surface convection to obtain $s \approx 0.7$, a result which has
been supported to some extent by a number of ground-based observations
of oscillations in solar-like stars. Very recently, Michel et
al. (2008) concluded from the first solar-like oscillation
observations made by CoRoT that intensity amplitudes were about 25\%
lower than the predictions of Samadi et al. (2005).

In this letter, we continue the work initiated by
Christensen-Dalsgaard \& Frandsen (1983) and Kjeldsen \& Bedding
(1995). However, rather than looking at the mode amplitudes we try
instead to find a simple scaling relation to describe the
\textit{lifetimes}, $\tau$, of the solar-like oscillations.  The
solar-like oscillations are damped (and excited) by convection, and so
the lifetimes are potentially an extremely useful diagnostic of
near-surface convection in stars. Our first motivation in seeking a
robust scaling relation is therefore to help elucidate the underlying
physics at play. Our second motivation is to enable simple, but
robust, lifetime predictions to be made for main-sequence, sub-giant
and red-giant stars. As explained below, the lifetimes affect the
detectability of the modes, so that lifetime predictions have a r\^ole
to play in target selection.

For a given mode amplitude, strong damping spreads the power of the
mode over a large range in frequency, implying a reduction in the
height of its peak in a power frequency spectrum compared to a mode
with weaker damping. With $\tau$ defined as the e-folding lifetime for
the amplitude, the mode peak observed in the frequency power spectrum
will have a FWHM linewidth of $\Delta = 1 /(\pi \tau)$.  If a peak is
well resolved, its height $H$ may be expressed in terms of the mode
amplitude, $A$ (either $v_{\rm osc}$ or $(\delta L/L)_{\lambda}$) and
mode linewidth, $\Delta$, according to
 \[
 H = \frac{2 A^2}{\pi \Delta} \propto \frac{A^2}{\Delta}.
 \]
Hence, when the mode is resolved, it is $H$ that determines the S/N in
power, not $A^2$ alone.

We note that a robust predictor is also desirable to fix the
underlying parameters of artificial asteroseismology data. Tests with
artificial data (e.g., the hare-and-hounds exercises conducted in
support of CoRoT and Kepler) are an important part of developing and
validating data analysis codes.

The layout of the rest of our paper is as follows. We begin in
Section~\ref{sec:data} with a description of the theoretically
computed mode lifetimes, and measured mode lifetimes, that were used
to infer the scaling relation. The measured mode lifetimes come from
observations of 12 main-sequence, sub-giant and red-giant stars, made
by ground-based telescopes and the CoRoT, WIRE and SMEI spacecraft.
We then proceed in Section~\ref{sec:res} to seek a relation between
mode lifetimes and the fundamental stellar parameters. Our main result
is that if we average lifetimes over the most prominent modes, the
average lifetime scales to a good approximation as $\left<\tau \right>
\propto T_{\rm eff}^{-4}$.  We finish in Section~\ref{sec:disc} with a
discussion of our scaling result, and its implications for future
selection of asteroseismic targets.

\section{Data}
\label{sec:data}

We consider mode lifetimes computed from the grid of stellar models
used by Chaplin et al. (2008) with measured mode lifetimes from 12
solar-like stars, sub-giants and giants.

The grid of stellar models used by Chaplin et al. had masses in the
range 0.7 to $1.3~M_{\odot}$, and ages in the range from the ZAMS to 9
Gyr. Padova isochrones (Bonatto et al. 2004; Girardi et al. 2002,
2004) were used to specify the primary characteristics of each model,
i.e., mass $M$, radius $R$, effective temperature $T_{\rm eff}$, and
luminosity $L$. The composition was fixed at $X=0.7$ and $Z=0.019$ for
all models. In order to estimate mode lifetimes for each model Chaplin
et al. (2008) then performed stellar equilibrium and pulsation
computations, as described in Balmforth (1992), Houdek et al. (1999),
and Chaplin et al. (2005). The pulsation computations required
estimates of $M$, $T_{\rm eff}$, $L$ and the composition as input. The
computations gave as output estimates of the powers and lifetimes of
the radial p modes of each stellar model. We used the powers and
lifetimes to compute the implied heights of the modes, and then
estimated an average lifetime $\left< \tau \right>$ for each model by
averaging lifetimes (and linewidths) over the five most prominent
radial modes, i.e., those with the largest heights.  More details may
be found in Chaplin et al. (2008).

We also use measurements of mode lifetimes of solar-like oscillations
from observations of 12 main-sequence, sub-giant and red-giant stars.
The measured lifetimes are shown in Table~\ref{tab:obs}, together with
estimates of stellar masses, luminosities, and effective temperatures,
the measured large frequency separations of the p-mode spectra, and
the list of references from which the data were taken.


\begin{table*}
	\begin{minipage}[t]{\textwidth}
		\caption{Mode lifetimes and fundamental stellar
		parameters for 12 solar-like stars, sub-giants and
		giants. }
		\label{tab:obs}
		\centering
		\renewcommand{\footnoterule}{}  
		\begin{tabular}{lccccc}
			\hline \hline
   			Name 				&	$M/M_{\sun}$			&	$L/L_{\sun}$			&	$T_{\mathrm{eff}}$			&	$\Delta \nu$					&	$\tau$		 				      	\\
   			\hline
   			$\xi$ Hydrae			&	3.07$^a$				&	$61.1 \pm 6.2^a$		&	$5000 \pm 100$ K$^a$		&	$7.11 \pm 6.14$ $\mu$Hz$^a$		&	2 days$^b$     				\\
   			$\nu$ Indi			 	&	0.847 $\pm$ 0.043$^c$	&	6.21 $\pm$ 0.23$^c$	&	5291 $\pm$ 34$ K$$^c$		&	24.25 $\pm$ 0.25  $\mu$Hz$^c$	&	16.2$_{-7.5}^{+44.6}$ days$^d$ 		\\
   			$\alpha$ Cen A		&	1.105 $\pm$ 0.007$^e$	&	1.522 $\pm$ 0.03$^e$	&	5810 $\pm$ 50 K$^e$		&	106.2$^f$						& $3.9 \pm 1.4$ days$^g$			\\
  		 	$\alpha$ Cen B		&	0.934 $\pm$ 0.006$^e$	&	0.503 $\pm$ 0.02$^e$	&	5260 $\pm$ 50 K$^e$		&	161.38 $\pm$ 0.06  $\mu$Hz$^h$	&	3.3$_{-0.9}^{+1.8}$ days$^h$			\\
   			$\beta$ Hydri			&	1.07 $\pm$ 0.03$^i$	&	3.51 $\pm$ 0.09$^i$	&	5872 $\pm$ 44 K$^i$		&	57.24 $\pm$ 0.16  $\mu$Hz$^j$	&	2.3$_{-0.5}^{+0.6}$ days$^j$ 		\\
  			HD49933				&	1.2$^k$				&	0.53 $\pm$ 0.01$^k$		&	6780 $\pm$ 130 K$^k$		&	85.9 $\pm$ 0.15  $\mu$Hz$^k$		&	2.3$_{-0.3}^{+0.5}$ days$^k$		\\
  			Procyon				&	1.48$^l$				&	6.9$^l$				&	6500 K$^l$				&	55.90 $\pm$ 0.08  $\mu$Hz$^l$	&	1.5$_{-0.8}^{+1.9}$ days$^m$			\\
  			$\tau$ Ceti			&	0.783 $\pm$ 0.012$^n$	&	0.488 $\pm$ 0.01$^n$	&	5264 K$^o$				&	 169 $\pm$ 0.2$\mu$Hz$^n$				&	1.7 $\pm$ 0.5 days$^n$				\\
  			$\varepsilon$ Oph		&	2.4 $\pm$ 0.4$^p$		&	59$^p$		&	4887 $\pm$ 100 K$^p$		&	5.3 $\pm$ 0.1  $\mu$Hz$^q$		&	2.7$_{-0.8}^{+0.6}$ days$^q$			\\
  			Arcturus				&	0.8 $\pm$ 0.3$^r$		&	200 $\pm$ 10$^r$		&	4290 $\pm$ 30 K$^r $		&	0.8 $\mu$Hz$^r$	&	24.1 days$^r$						\\
  			$\beta$ Ursae Minoris	&	1.3 $\pm$ 0.3$^s$		&      475 $\pm$ 30$^s$		&	 4040 $\pm$ 100 K$^s$		&	0.7 $\pm$ 0.1  $\mu$Hz$^s$		&	18 $\pm$ 9 days$^s$		\\
  			Sun					&	1					&	1					&	5780	 K					&	134 $\mu$Hz					& 	3.2 $\pm$ 0.2 days$^t$				\\
			\hline
		\end{tabular}
	\end{minipage}
	\vspace{0.2cm}
		\endnote{References: $^a$~Frandsen et al. (2002);
		$^b$~Stello et al. (2006); $^c$~Bedding et al. (2006);
		$^d$~Carrier et al. (2007); $^e$~Miglio \&
		Montalb{\'a}n (2005); $^f$~Bedding et al.(2004); $^g$~Fletcher et al. (2006);
		$^h$~Kjeldsen et al. (2005); $^i$~North et al. (2007);
		$^j$~Bedding et al. (2007); $^k$~Appourchaux et
		al. (2008); $^l$~Leccia et al. (2007); $^m$~Arentoft et al. (2008); $^n$~Teixeira et al. (2008);
		$^o$~Pijpers (2003); $^p$~de Ridder et al. (2006);
		$^q$~Barban et al. (2007); $^r$~Tarrant et al. (2007);
		$^s$~Tarrant et al. (2008); $^t$~Chaplin et al. (2005).}
\end{table*}


It is important to point out that the mode lifetimes were not measured
in the same way on all 12 stars, and the quoted averages are not
entirely self consistent; nor do the quoted values necessarily
correspond precisely to an average over the five strongest radial
modes (the datum chosen for the model data). However, given the level
of precision in the measurements, and given that, in many cases, it
was only possible to estimate the lifetimes of a small number of modes
in the vicinity of the highest amplitude, we feel this is not a major
cause for concern.

The mismatch between the measured lifetimes of the different stars
will no doubt introduce some extra scatter into the lifetime-stellar
parameter relations we seek to constrain. We also note that for the
stars observed by the Kjeldsen \& Bedding group ($\nu$~Indi,
$\alpha$~Cen~B, $\beta$~Hydri, Procyon and $\tau$~Ceti), lifetimes
were evaluated from frequency scatter of mode frequency ridges in the
Echelle diagram without including the effects of rotation, which are
known to artificially lower the value of the mode lifetime (see Bazot
et al. 2007 for discussion). We comment in Section~\ref{sec:res} below
on the impact of this bias on our results.

\section{Results}
\label{sec:res}


\begin{figure*}
 \centerline {\epsfxsize=8.5cm\epsfbox{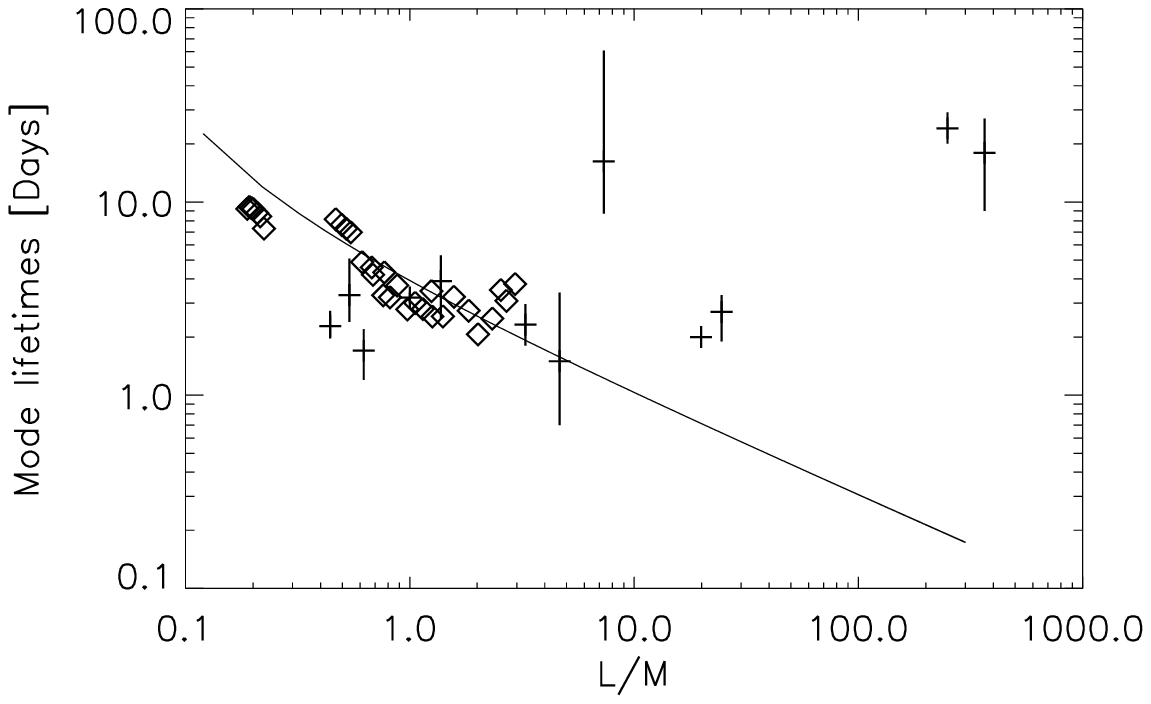}
              \epsfxsize=8.5cm\epsfbox{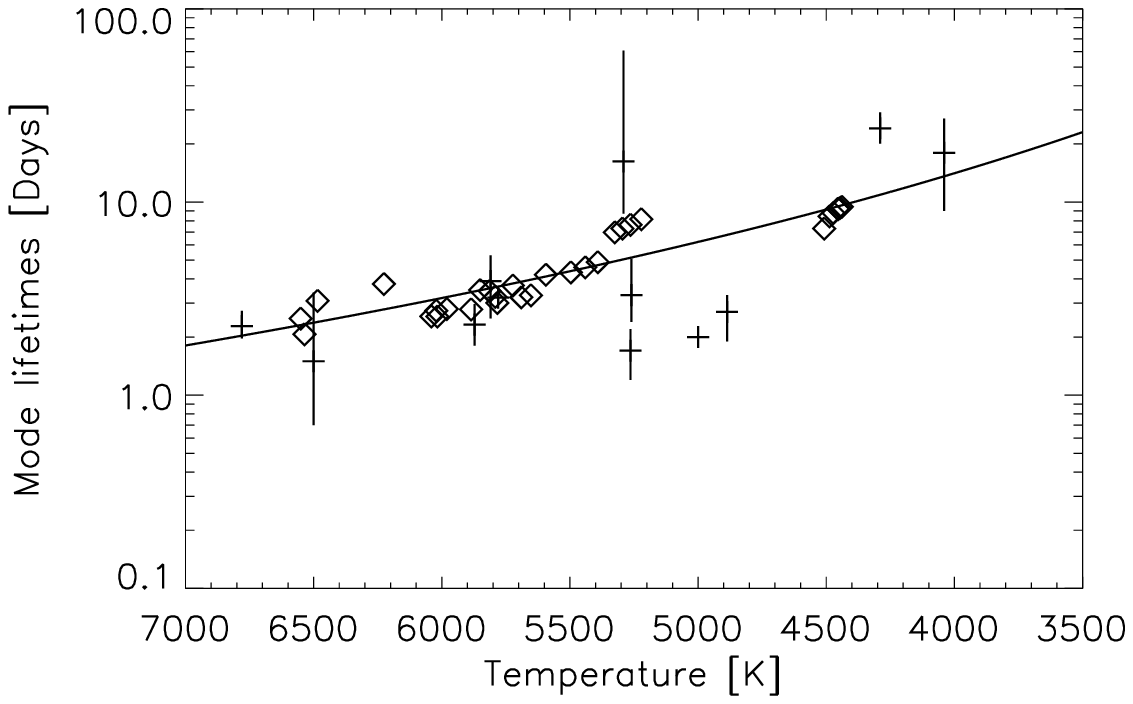}}

 \caption{Average mode lifetimes $\left< \tau \right>$ from the stellar
 equilibrium and pulsation computations (diamonds) and from
 observations of 12 stars (crosses), plotted as a function of: the
 $L/M$ of each star (left-hand panel); and the effective temperature,
 $T_{\rm eff}$, of each star.}

 \label{fig:res}
\end{figure*}


Following Kjeldsen \& Bedding (1995), we first sought to obtain a
power-law relation between the average lifetimes and $L/M$. The
left-hand panel of Fig.~\ref{fig:res} plots, as a function of $L/M$,
the average lifetimes from the stellar equilibrium and pulsation
computations (diamond symbols), and the average lifetimes from the
observations (crosses with error bars). The average lifetimes from the
pulsation computations -- which we recall do not include evolved stars
-- follow a well-constrained power-law relation of the form
 \begin{equation}
 \left< \tau \right> \propto \left( L/M \right)^{-0.51 \pm 0.05}.
 \end{equation}
The solid line shows the best-fitting power law.  The average
lifetimes of the observed main-sequence stars show a rough match to
this power law; however, the more evolved stars with large $L/M$
depart significantly from the prediction. We conclude that a power law
in $L/M$ fails to provide an adequate scaling relation for
main-sequence, sub-giant and giant stars.

It struck us that in order to get the observed lifetimes to follow a
single curve, we needed an independent variable which reversed the
order on the abscissa of the main-sequence points and giant
points. Use of $T_{\rm eff}$ as the independent variable would fulfill
this requirement. The right-hand panel of Fig.~\ref{fig:res} therefore
plots the lifetime data as a function of $T_{\rm eff}$. The pulsation
computations again follow a well-constrained power law, described by
 \begin{equation}
 \left< \tau \right> \propto T_{\rm eff}^{-3.7 \pm 0.3},
 \end{equation}
but this time the observational data on the main-sequence stars,
sub-giants \emph{and} giants also show a good match to the
best-fitting curve. We add that the model prediction for $\xi$~Hydrae
presented by Houdek \& Gough (2002) is not inconsistent with this
power law. However, Houdek \& Gough did make different assumptions
about the ``non-locality'' of the convection to those adopted in the
models used here. We are currently working on new models for a
selection of red giant stars, which will provide a consistent
treatment of the convection parameters.

We therefore propose that a power law of the form
 \[
 \left< \tau \right> \propto T_{\rm eff}^{-4}
 \]
does provide an adequate scaling relation for main-sequence, sub-giant
and giant stars.

Some of the measured lifetimes, close to $T_{\rm eff} \sim 5000\,\rm
K$, are noticeably shorter than the lifetimes implied by the
best-fitting power law. These measurements were made by the Kjeldsen
\& Bedding group, and we suggest that the results for these stars may
have been affected by not including rotation in the analysis
(cf. Section~\ref{sec:data}).

The result we obtain when fitting the lifetimes from the stellar
equilibrium and pulsation computations to a power law in $L/M$
disagrees somewhat with the result of Chaplin et al. (2007). Chaplin
et al. obtained $\Delta \propto \left(L/M \right)^{1.6 \pm 0.3}$,
while our result here implies $\Delta \propto \left( L/M
\right)^{0.5}$. Chaplin et al. used pulsation computations of a sample
of models matching 22 solar-like stars in the Mount Wilson Ca~{\sc
ii}~H and K survey (Baliunas et al. 1995). All models had $L/M$ lower
than $\sim 1.3$-times the solar value. We suggest, with reference to
Fig.~\ref{fig:res}, that this is a too small range in $L/M$ to extract
a robust relation.

\section{Discussion}
\label{sec:disc}

The aim of this study was to find a simple scaling relation between
the mean lifetimes, $\left< \tau \right>$, of the most prominent
solar-like p modes in stars, and the fundamental stellar
parameters. We used predicted lifetimes from stellar equilibrium and
pulsation computations, and measured mode lifetimes from observations
of 12 main-sequence, sub-giant and red-giant stars. Our study suggests
that the lifetimes follow to a good approximation a scaling relation
that depends only on effective temperature:
 \begin{equation}
 \left< \tau \right> \propto T_{\rm eff}^{-4}.
 \end{equation}
The fourth-power dependence on effective temperature deserves some
comment. The total surface flux radiated by a star is of course by
definition proportional to $T_{\rm eff}^4$, and we would expect that,
to first order, this will also correspond approximately to the surface
convective heat flux in stars with near-surface convection zones. The
total heat flux transported by the convection is clearly an important
parameter in fixing the strength of damping of solar-like
oscillations. It is perhaps worth noting (see below) that a similar,
simple dependence on effective temperature is not seen in the
\emph{amplitudes} of the solar-like p modes.

We may combine our scaling relation for average lifetime with the
scaling relation of Kjeldsen \& Bedding (1995) for the maximum mode
amplitude, $A$, to yield a scaling relation for the maximum mode
height, $H$, in the frequency power spectrum. For narrow-band
intensity observations, Kjeldsen \& Bedding proposed:
 \[
 A = (\delta L/L)_{\lambda} \propto 
     \frac{L}{M T_{\rm eff}^2} \propto \frac{T_{\rm eff}^2}{g}.
 \]
Combining with our prediction for lifetime, we obtain
 \begin{equation}
 H \propto g^{-2}.
 \label{eq:h}
 \end{equation}
We predict that the maximum mode height -- which fixes the S/N, and
hence the visibility, of peaks when they are resolved in the frequency
power spectrum -- depends \emph{predominantly} on the surface gravity
of stars, when the observations are made in intensity. We use the
qualification in the preceding sentence because the above does assume
that the Kjeldsen \& Bedding scaling for $(\delta L/L)_{\lambda}$ is
accurate. While this scaling is supported to a large extent by
observations, there is evidence for some departures from the scaling,
most notably for hotter solar-like stars (as noted in
Section~\ref{sec:intro} above). When observations are instead made in
Doppler velocity -- which, using Kjeldsen \& Bedding's proposed
scaling, implies $A \propto L/M$ -- we find
 \begin{equation}
 H \propto \frac{T_{\rm eff}^4}{g^2}.
 \label{eq:h1}
 \end{equation}

Since the surface gravity changes fairly slowly along the main
sequence, our new scaling relation for $H$ given by
Equation~\ref{eq:h}, which assumes narrow-band intensity observations,
suggests that stars notably cooler than the Sun might have mode
heights that are in fact comparable to those of solar-like stars that
are hotter than the Sun.  Previously, asteroseismic target selections
were based only on the expected amplitudes, mainly because we had no
idea about how the mode lifetimes scaled with fundamental stellar
parameters. This would lead one to favour hotter stars as targets,
while our result here suggests cooler solar-like stars may actually be
very good targets as well.  When observations are instead made in
Doppler velocity (Equation~\ref{eq:h1}), our results imply that $H$
increases with increasing effective temperature.

Our prediction that cooler stars show longer lifetimes brings another
added benefit: it is easier to estimate stellar rotation rates using
asteroseismology, because the ratio between the rotational splittings
and the mode linewidths will generally be higher than in hotter stars
(see also Chaplin et al. 2008). This is important, because one of the
main goals of asteroseismology is to measure surface and internal
differential rotation (see also: Gizon \& Solanki 2003, 2004; and
Ballot et al. 2006).

We therefore suggest that in future our scaling relations for mode
height (Equations~\ref{eq:h} and~\ref{eq:h1}), and not the previously
used relations for mode amplitude, be used when selecting targets for
asteroseismology.

\begin{acknowledgements}

WJC, GH and YE acknowledge the support of the UK Science and
Technologies Facilities Council (STFC). CK acknowledges financial
support from the Danish Natural Sciences Research Council. The authors
would like to thank D.~Gough for useful discussions. This paper was
inspired by work conducted as part of the
asteroFLAG\footnote{http://www.issibern.ch/teams/Astflag} project. We
acknowledge the International Space Science Institute (ISSI), which
provides support for asteroFLAG.  This work was also supported by the
European Helio- and Asteroseismology Network
(HELAS)\footnote{http://www.helas-eu.org}, a major international
collaboration funded by the European Commission's Sixth Framework
Programme.

\end{acknowledgements}


\begin{thebibliography}{}

\bibitem[Appourchaux et al.(2008)]{2008A&A...488..705A} Appourchaux,
T., et al.\ 2008, \aap, 488, 705

\bibitem[Arentoft et al.(2008)]{2008ApJ...687.1180A} Arentoft, T., et
al.\ 2008, \apj, 687, 1180

\bibitem[Baliunas et al.(1995)]{1995ApJ...438..269B} Baliunas, S.~L.,
et al.\ 1995, \apj, 438, 269

\bibitem[Ballot et al.(2006)]{2006MNRAS.369.1281B} Ballot, J.,
Garc{\'{\i}}a, R.~A., \& Lambert, P.\ 2006, \mnras, 369, 1281

\bibitem[Balmforth(1992)]{1992MNRAS.255..603B} Balmforth, N.~J.\ 1992,
\mnras, 255, 603

\bibitem[Barban et al.(2007)]{2007A&A...468.1033B} Barban, C., et al.\
2007, \aap, 468, 1033

\bibitem[Basri et al.(2005)]{2005NewAR..49..478B} Basri, G., Borucki,
W.~J., \& Koch, D.\ 2005, New Astronomy Review, 49, 478

\bibitem[Bazot et al.(2007)]{2007A&A...470..295B} Bazot, M., Bouchy,
F., Kjeldsen, H., Charpinet, S., Laymand, M., \& Vauclair, S.\ 2007,
\aap, 470, 295

\bibitem[Bedding et al.(2004)]{2004ApJ...614..380B} Bedding, T.~R.,
Kjeldsen, H., Butler, R.~P., McCarthy, C., Marcy, G.~W., O'Toole,
S.~J., Tinney, C.~G., \& Wright, J.~T.\ 2004, \apj, 614, 380

\bibitem[Bedding et al.(2006)]{2006ApJ...647..558B} Bedding, T.~R., et
al.\ 2006, \apj, 647, 558

\bibitem[Bedding et al.(2007)]{2007ApJ...663.1315B} Bedding, T.~R., et
al.\ 2007, \apj, 663, 1315

\bibitem[Bonatto et al.(2004)]{2004A&A...415..571B} Bonatto, C., Bica,
E., \& Girardi, L.\ 2004, \aap, 415, 571

\bibitem[Bruntt(2007)]{bruntt07} Bruntt H., 2007, CoAst, 150, 326

\bibitem[Carrier et al.(2007)]{2007A&A...470.1059C} Carrier, F., et
al.\ 2007, \aap, 470, 1059

\bibitem[Chaplin et al.(2008)]{2008A&A...485..813C} Chaplin, W.~J.,
Houdek, G., Appourchaux, T., Elsworth, Y., New, R., \& Toutain, T.\
2008, \aap, 485, 813


\bibitem[Christensen-Dalsgaard \& Frandsen(1983)]{1983SoPh...82..469C}
Christensen-Dalsgaard, J., \& Frandsen, S.\ 1983, \solphys, 82, 469

\bibitem[Christensen-Dalsgaard et al.(2007)]{} Christensen-Dalsgaard
J., Arentoft T., Brown T. M., Gilliland R. L., Kjeldsen H., Borucki
W. J., Koch D., 2007, CoAst, 150, 350

\bibitem[Fletcher et al.(2006)]{} Fletcher S. T., Chaplin W. J.,
Elsworth Y., Schou J., Buzasi D., 2006, \mnras, 371, 935

\bibitem[Frandsen et al.(2002)]{2002A&A...394L...5F} Frandsen, S., et
al.\ 2002, \aap, 394, L

\bibitem[Girardi et al.(2002)]{2002A&A...391..195G} Girardi, L.,
Bertelli, G., Bressan, A., Chiosi, C., Groenewegen, M.~A.~T., Marigo,
P., Salasnich, B., \& Weiss, A.\ 2002, \aap, 391, 195

\bibitem[Girardi et al.(2004)]{2004A&A...422..205G} Girardi, L.,
Grebel, E.~K., Odenkirchen, M., \& Chiosi, C.\ 2004, \aap, 422, 20

\bibitem[Gizon \& Solanki(2003)]{2003ApJ...589.1009G} Gizon, L., \&
Solanki, S.~K.\ 2003, \apj, 589, 1009

\bibitem[Gizon \& Solanki(2004)]{2004SoPh..220..169G} Gizon, L., \&
Solanki, S.~K.\ 2004, \solphys, 220, 169

\bibitem[Houdek et al.(1999)]{1999A&A...351..582H} Houdek, G.,
Balmforth, N.~J., Christensen-Dalsgaard, J., \& Gough, D.~O.\ 1999,
\aap, 351, 582

\bibitem[Houdek \& Gough(2002)]{2002MNRAS.336L..65H} Houdek, G., \&
Gough, D.~O.\ 2002, \mnras, 336, L65

\bibitem[Kjeldsen \& Bedding(1995)]{1995A&A...293...87K} Kjeldsen, H.,
\& Bedding, T.~R.\ 1995, \aap, 293, 87

\bibitem[Kjeldsen et al.(2005)]{2005ApJ...635.1281K} Kjeldsen, H., et
al.\ 2005, \apj, 635, 1281

\bibitem[Leccia et al.(2007)]{2007A&A...464.1059L} Leccia, S.,
Kjeldsen, H., Bonanno, A., Claudi, R.~U., Ventura, R., \& Patern{\`o},
L.\ 2007, \aap, 464, 1059

\bibitem[Matthews et al.(2007)]{2007CoAst...350.353M} Matthews~J., et
al., 2007, CoAst, 350, 333

\bibitem[Michel et al.(2008)]{2008Sci...322..558M} Michel, E., et al.\
2008, Science, 322, 558

\bibitem[Miglio \& Montalb{\'a}n(2005)]{2005A&A...441..615M} Miglio,
A., \& Montalb{\'a}n, J.\ 2005, \aap, 441, 615

\bibitem[North et al.(2007)]{2007MNRAS.380L..80N} North, J.~R., et
al.\ 2007, \mnras, 380, L80

\bibitem[Pijpers(2003)]{2003A&A...400..241P} Pijpers, F.~P.\ 2003,
\aap, 400, 241

\bibitem[Retter et al.(2003)]{2003ApJ...591L.151R} Retter, A.,
Bedding, T.~R., Buzasi, D.~L., Kjeldsen, H., \& Kiss, L.~L.\ 2003,
\apjl, 591, L151

\bibitem[de Ridder et al.(2006)]{2006A&A...448..689D} de Ridder, J.,
Barban, C., Carrier, F., Mazumdar, A., Eggenberger, P., Aerts, C.,
Deruyter, S., \& Vanautgaerden, J.\ 2006, \aap, 448, 689

\bibitem[Samadi et al.(2005)]{2005JApA...26..171S} Samadi, R., Goupil,
M.-J., Alecian, E., Baudin, F., Georgobiani, D., Trampedach, R.,
Stein, R., \& Nordlund, {\AA}.\ 2005, Journal of Astrophysics and
Astronomy, 26, 171

\bibitem[Stello et al.(2006)]{2006A&A...448..709S} Stello, D.,
Kjeldsen, H., Bedding, T.~R., \& Buzasi, D.\ 2006, \aap, 448, 709

\bibitem[Tarrant et al.(2007)]{2007MNRAS.382L..48T} Tarrant, N.~J.,
Chaplin, W.~J., Elsworth, Y., Spreckley, S.~A., \& Stevens, I.~R.\
2007, \mnras, 382, L48

\bibitem[Tarrant et al.(2008)]{2008A&A...483L..43T} Tarrant, N.~J.,
Chaplin, W.~J., Elsworth, Y., Spreckley, S.~A., \& Stevens, I.~R.\
2008, \aap, 483, L43

\bibitem[Teixeira et al.(2008)]{2008arXiv0811.3989T} Teixeira, T.~C.,
et al.\ 2008, \apj, accepted, arXiv:0811.3989

\end{thebibliography}
\end{document}